# Vortex states in a mesoscopic superconducting triangle

V.R. Misko [a,1], V.M. Fomin [a,2], J.T. Devreese [a,3,4], V.V. Moshchalkov [b]

[a]*Theoretische Fysica van de Vaste Stof, Universiteit Antwerpen (UIA), Universiteitsplein 1, B-2610 Antwerpen, Belgium*
[b]*Laboratorium voor Vaste Stoffysica en Magnetisme, Katholieke Universiteit Leuven, Celestijnenlaan 200 D, B-3001 Leuven, Belgium*
[1]*Department of Theory of Semiconductors and Quantum Electronics, Institute of Applied Physics, Academy of Sciences of Moldova, str. Academiei 5, MD-2028 Kishinev, Republic of Moldova*
[2]*Department of Theoretical Physics, State University of Moldova, str. Mateevici 60, MD-2009 Kishinev, Republic of Moldova*
[3]*Universiteit Antwerpen (RUCA), Groenenborgerlaan 171, B-2020 Antwerpen, Belgium*
[4]*Technishche Universiteit Eindhoven, P.O. Box 513, 5600 MB Eindhoven, The Netherlands*

The set of the nonlinear Ginzburg-Landau equations is solved for an Al mesoscopic superconducting triangle of finite thickness. We calculate the distributions of the superconducting phase in the triangle and of the magnetic field in and near the triangle. The distribution of the superconducting phase in the triangle is studied as a function of the applied magnetic field. Possible scenarios of penetration of the magnetic field into the triangle are analyzed. We consider two different states: a single vortex state and a state in the form of a symmetric combination of three vortices and an antivortex with vorticity $L_a = -2$ ("3 – 2" combination). The free energy calculations show that a single vortex penetrates the triangle through a midpoint of one side. The "3 – 2" combination turns out to be thermodynamically preferable when the vortices are close to the center of the triangle. Equilibrium is achieved when a single vortex (or each component of the "3 – 2" combination) is in the center of the triangle.

## I. Introduction

Technological progress in fabrication of nanosized superconductors with sharp corners has stimulated investigations of mesoscopic structures such as wedges,[1] squares and square loops,[2,3] triangles,[4,5] etc. The critical parameters (critical magnetic field, critical current) are enhanced in these structures as compared with bulk materials.

Along with the enhancement of the critical parameters, penetration of magnetic field into a mesoscopic structure with sharp corners is strongly modified in comparison with macroscopic samples. In particular, a symmetry imposed by the geometrical shape of a mesoscopic superconductor interferes with the trigonal symmetry imposed by the interaction between vortices.[6] Based on the linearized Ginzburg-Landau (GL) equations and on the assumption of a constant magnetic field, it was shown in Refs. [4], that in triangles and squares, respectively, $C_3$- and $C_4$- symmetric configurations of vortices and antivortices may be realized in the close vicinity to the phase boundary. In Ref. [5], for squares and triangles with the thickness $d = 0.1\varkappa$ and with the GL parameter $\varkappa = 0.28$, regions of magnetic field were found, when a giant vortex state coexisted with several separated vortices. However, no multivortices or antivortices were revealed in Ref. [5] near the phase boundary. In the present paper, on the basis of a full set of the *nonlinear* GL equations, we study the magnetic field distributions in and near a mesoscopic triangle and vortex states in the triangle. The evolution of the superconducting state with no vortex in the triangle is analyzed when the applied magnetic field $H_0$ increases. Possible scenarios are discussed, which show how a vortex penetrates into a triangle.

## II. The Ginzburg-Landau equations in a mesoscopic triangle

To describe the superconducting state in a mesoscopic triangle, we use the GL equations for the order parameter $\psi$ and the vector potential $\mathbf{A}$ of a magnetic field $\mathbf{H}=\nabla\times\mathbf{A}$[7]

$$\frac{1}{2m}\left(-i\hbar\nabla-\frac{2e}{c}\mathbf{A}\right)^2\psi+a\psi+b|\psi|^2\psi=0$$

$$\nabla^2\mathbf{A}=\frac{4\pi i e\hbar}{mc}(\psi^*\nabla\psi-\psi\nabla\psi^*)+\frac{16\pi e^2}{mc^2}\mathbf{A}|\psi|^2 \qquad (1)$$

with the boundary condition

$$\mathbf{n}\cdot\left(-i\hbar\nabla\psi-\frac{2e}{c}\mathbf{A}\psi\right)=0, \qquad (2)$$

where $\mathbf{n}$ is the unit vector normal to the boundary, and $a$ and $b$ are the coefficients of the GL theory.

There are various ways to transform the GL equations to a dimensionless form. If the temperature-dependent coherence length $\xi(T)$ is chosen as the unit of length, the GL equations take a relatively simple form, but the sizes of a sample become a function of temperature. To avoid this inconvenience, we use in this paper the coherence length at zero temperature $\xi(0)$ as the unit of length.

As a result, the dimensionless GL equations take the form:

$$(-i\nabla-\mathbf{A})^2\psi-\psi\left[\left(1-\frac{T}{T_c}\right)-|\psi|^2\right]=0,$$

$$\kappa^2\Delta\mathbf{A}=\frac{i}{2}(\psi^*\nabla\psi-\psi\nabla\psi^*)+\mathbf{A}|\psi|^2, \qquad (3)$$

$$\mathbf{n}\cdot(-i\nabla\psi-\mathbf{A}\psi)\big|_{boundary}=0, \qquad (4)$$

where $\kappa=\lambda(T)/\xi(T)$, and $T_c$ is the critical temperature.

The boundary condition for the vector potential

$$\mathbf{A}\big|_{boundary\ sr}=\left(-\frac{H_0 y}{2},\frac{H_0 x}{2},0\right) \qquad (5)$$

is determined at the boundary of the *simulation region ("sr")*, which is chosen large enough to ensure that all changes of the magnetic field occur *inside* this region.[8–10]

A mesoscopic superconducting triangle is supposed to have a thickness $d\sim 20$ nm that is much less than $\xi(T)$ [$\xi(T=0.96T_c)=500$ nm]. Therefore, the order parameter can be considered uniform in the $z$-direction. This approximation was used in Refs. [5, 8, 9] for the description of thin mesoscopic samples with $d\ll\xi(T)$. On the contrary, in the case of a mesoscopic bridge, all the quantities entering the GL equations depend on $z$, and we have to deal actually with a three-dimensional problem.[10]

### III. The vortex state in the triangle: profiles of the magnetic field

The solutions of the GL equations are obtained for the effective value of the GL parameter $k$ = 1.2 (with decreasing dimensions of a mesoscopic sample, the GL parameter $k$ increases, and metals effectively exhibit properties of type II superconductors[2,3,6]). The calculated distributions of the magnetic field are plotted for $L = 0$ (no vortex), $T/T_c = 0.9$, $H_0 = 0.8H_c(0)$ (Fig.1a) and for $L = 1$ (a single vortex), $T/T_c = 0.9$, $H_0 = 1.4H_c(0)$ (Fig.1b). The distributions are represented by profiles along different cross-sections through the triangle. The magnetic field remains equal to $H_0$ up to a distance of about 300 nm from the triangle and is *enhanced* (see Fig. 1) near the boundary of the triangle. This enhancement is due to the demagnetization effect: the magnetic field is partially expelled from the triangle. Presence of a vortex in the triangle results in more complicated profiles (Fig. 1b) than those for the case $L = 0$. For the state with a vortex, along with the features described above for the state with $L = 0$, there appears an additional maximum of the magnetic field, which is related to a vortex.

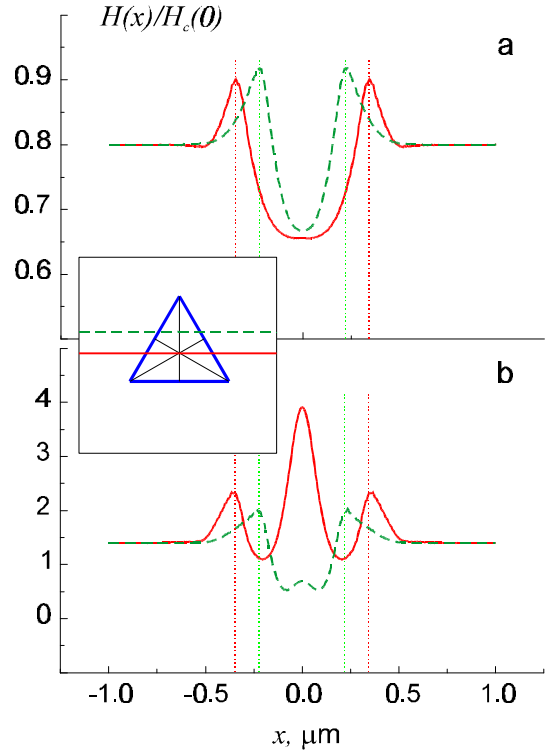

Fig. 1. Profiles of the magnetic field $H(x)$ along various cross-sections (see inset) of the equilateral triangle with side $a = 1\mu m$ for $L = 0$ (a), $L = 1$(b).

### IV. Penetration of a vortex into the triangle: possible scenarios

When the applied magnetic field reaches some critical value $H_{c1}^{tr}$, a vortex (or vortices) starts to penetrate the triangle. The critical field $H_{c1}^{tr}$ is a function of temperature and is enhanced in the mesoscopic triangle, as compared with the bulk first critical field $H_{c1}$. In order to analyze possible scenarios of penetration of a vortex into the triangle, we first consider evolution of the order parameter distribution with increasing $H_0$ for $H_0 < H_{c1}^{tr}$. Calculations of the free energy show that the state with $L = 0$ is thermodynamically favorable up to $H_0 = 1.5H_c(0)$ at $T/T_c = 0.96$ (Fig. 2).

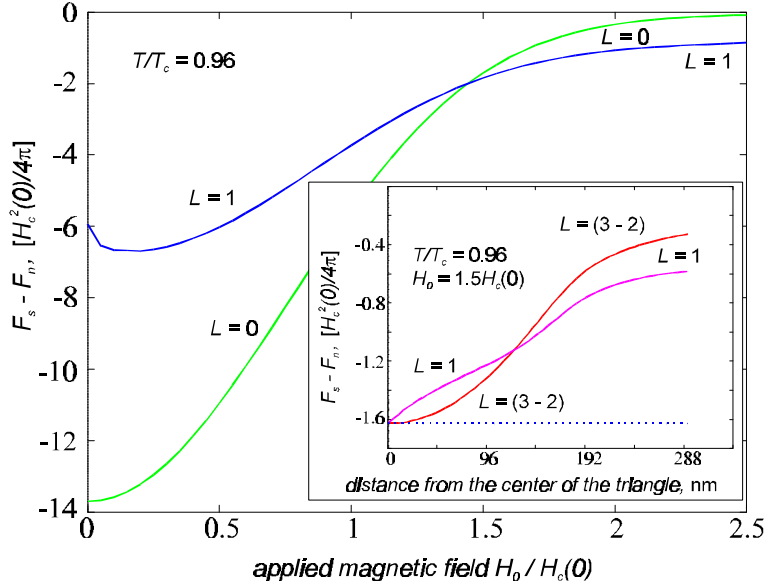

Fig. 2. The free energy [measured in $H_c^2(0)/4\pi$] of the superconducting state in the mesoscopic triangle, as a function of the applied magnetic field, for $L = 0$ and $L = 1$. Inset: the free energy is plotted for the set of stationary configurations, which are characterized by different displacements of a vortex from the center of the triangle (for a single vortex, $L = 1$) or of three vortices [for a symmetric combination of three vortices and an ativortex with vorticity $L_a = -2$, $L = (3-2)$].

The squared modulus of the order parameter $|\psi(x,y)|^2$ is shown in Fig. 3 as a function of the applied magnetic field for $T/T_c = 0.96$. With increasing the applied magnetic field, there appear minima of $|\psi(x,y)|^2$ at the midpoints of the sides of the triangle. The $|\psi(x,y)|^2$ distributions exhibit the $C_3$-symmetry. Obviously, a vortex (or vortices) will penetrate the triangle through midpoints of the sides of the triangle. Let us consider two possible scenarios of penetration of a vortex into the triangle: (i) a *single vortex* enters the triangle through a midpoint of one of its sides (Fig.4a); (ii) a $C_3$-symmetric *combination* consisting of *three vortices* and an *antivortex* with vorticity $L_a = -2$ appears in the triangle (Fig. 4e). This combination of vortices and an antivortex with the total vorticity $L = (3 - 2) = 1$ will be referred to as a "3 − 2" combination. It is noteworthy, that the symmetry of the order parameter distribution before a vortex penetration holds after the "3 − 2" combination appears in the triangle. However, our calculations of the free energy show (inset to Fig. 2) that penetration of a single vortex through a midpoint of one side of the triangle is energetically more favorable than appearance of the "3 − 2" combination. At the same time, as it follows from Fig. 2 [in particular, for $T/T_c = 0.96$ and $H_0 = 1.5H_c(0)$], the "3 − 2" combination turns out to be thermodynamically preferable when the vortices are close to the center of the triangle. This result can be understood if we consider the distribution of the order parameter when a vortex (or each component of the "3 − 2" combination) is positioned at the center of the triangle. A small displacement of the single vortex from the center of the triangle leads to a distortion of the $C_3$-symmetric function $|\psi(x,y)|^2$. This distortion affects the regions, where $|\psi(x,y)|^2$ has relatively high values, what leads to an increase of the free energy. Contrary to the above picture, a small

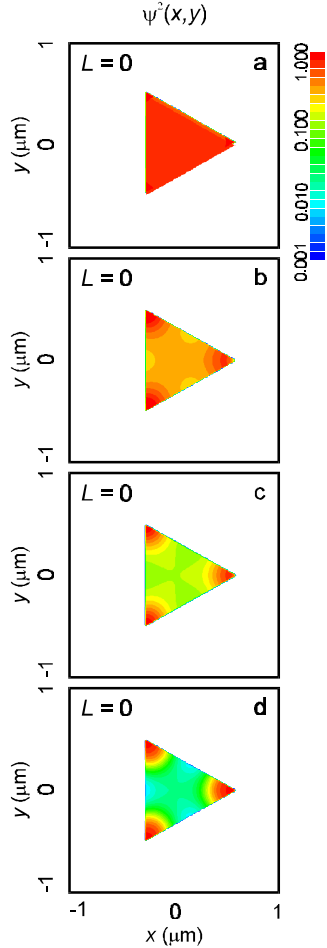

Fig. 3. The distribution of $|\psi(x,y)|^2$ as a function of the applied magnetic field: $H_0=0$ (a), $H_0=0.45H_c(0)$ (b), $H_0 = 0.8H_c(0)$ (c), $H_0 =1.5H_c(0)$ (d), for vorticity $L = 0$ (there is no vortex in the triangle) and for temperature $T/T_c = 0.96$.

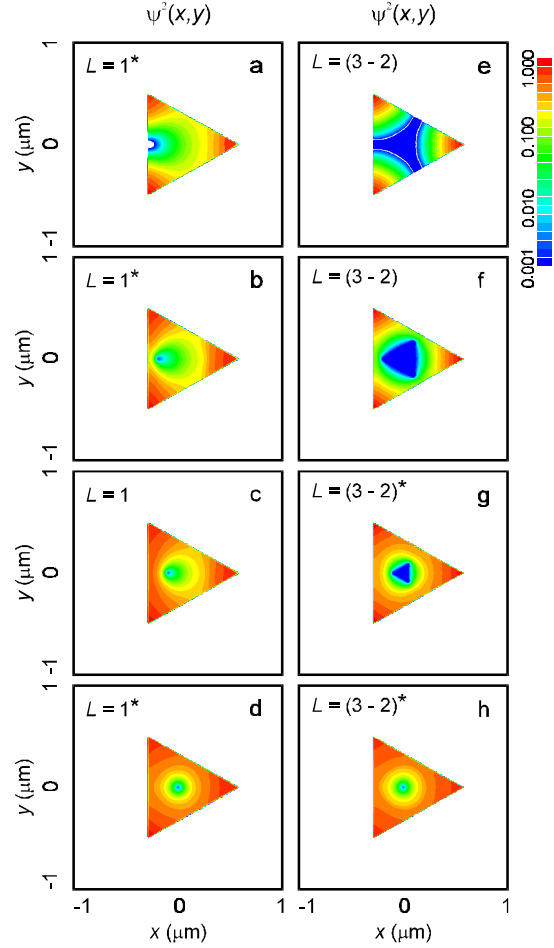

Fig. 4. Penetration of a vortex into the triangle: $|\psi(x,y)|^2$ as a function of the displacement of a single vortex $d_v$ from the center of the triangle for $L = 1$ (a-d), or as a function of displacement of three vortices $d_{3v}$ (an antivortex with vorticity $L_a = -2$ is placed in the center of the triangle) for $L = (3 - 2)$ (e-h). The displacements are: $d_v = d_{3v} = 288$ nm (a, e), 192 nm (b, f), 96 nm (c, g), 0 (d, h). When a vortex (vortices) is far from the center of the triangle ($d_v = d_{3v} = 288$ nm, 192 nm), the states with $L = 1$ (a, b) are characterized by a lower free energy (such states are marked with a star) than the corresponding states with $L = (3 - 2)$ (e, f). When a vortex (vortices) is close to the center of the triangle ($d_v = d_{3v} = 96$ nm), the state with $L = (3 - 2)$ (g) has a lower free energy than the corresponding state with $L = 1$(c). For $d_v = d_{3v} = 0$, the states with $L = 1$ (d) and with $L = (3 - 2)$ (h) coincide.

displacement of the vortices of the "3 − 2" combination means a small increase of size of the normal region near the center of the triangle. This small increase does not appreciably affect the regions characterized by high values of $|\psi(x,y)|^2$ and, therefore, is not accompanied by a noticeable increase in the free energy. As a result, in the vicinity of the center of the triangle, a

prepared state in the form of the symmetric combination of three vortices and antivortex with vorticity $L_a = -2$ is the thermodynamically preferable state. Our analysis shows also that when approaching the phase boundary, a difference between the free energy of the state with a single vortex entering the triangle and that of the combination "3 − 2" vanishes.

## V. Conclusions

The distribution of the superconducting phase in a triangle of finite thickness and the distribution of the magnetic field in and near the triangle are analyzed on the basis of the numerical solution of the nonlinear GL equations. The magnetic field is enhanced near the boundary of the triangle. For $L = 0$, the distributions of the superconducting phase exhibit the $C_3$-symmetry. Penetration of the magnetic field into the triangle is investigated. The free energy analysis shows that: *(i)* a single vortex enters the triangle through a midpoint of one of its sides, *(ii)* a $C_3$-symmetric combination of three vortices and an antivortex with vorticity $L_a = -2$ turns out to be energetically preferable when the vortices are close to the center of the triangle.

## Acknowledgments

This work has been supported by GOA BOF UA 2000, IUAP, the FWO-V projects Nos. G.0287.95, G.0306.00, G.0274.01, WOG WO.025.99N (Belgium), and the ESF Programme VORTEX.

## References

[1] V. M. Fomin, J. T. Devreese and V. V. Moshchalkov, Europhys. Lett. **42**, 553 (1998); **46**, 118 (1999).
[2] V. V. Moshchalkov, L. Gielen, C. Strunk, R. Jonckheere, X. Qiu, C. Van Haesendonck, and Y. Bruynseraede, Nature (London) **373**, 319 (1995).
[3] V. M. Fomin, V. R. Misko, J. T. Devreese, V. V. Moshchalkov, Solid State Commun. **101**, 303 (1997); Phys. Rev. B **58**, 11703 (1998).
[4] L. F. Chibotaru, A. Ceulemans, V. Bruyndoncx and V. V. Moshchalkov, Nature (London) **408**, 833 (2000); Phys. Rev. Lett. **86**, 1323 (2001).
[5] B. J. Baelus and F. M. Peeters, cond-mat/0106601.
[6] M. Tinkham, *Introduction to Superconductivity*, 2-nd Ed., McGraw-Hill, New-York, 1996.
[7] V. L. Ginzburg and L. D. Landau, Zh. Eksp. Teor. Fiz. **20**, 1064 (1950).
[8] P. Singha Deo, V. A. Schweigert, F. M. Peeters, and A. K. Geim, Phys. Rev. Lett. **79**, 4653 (1997).
[9] V. A. Schweigert and F. M. Peeters, Phys. Rev. B **57**, 13817 (1998).
[10] V. R. Misko, V. M. Fomin, J. T. Devreese, Phys. Rev. B **64**, 014517 (2001).